\documentclass[12pt]{iopart}
\usepackage{iopams}
\usepackage{setstack}

\newcommand{\half}{\mbox{$\textstyle\frac{1}{2}$}}
\newcommand{\cP}{{\cal P}}
\newcommand{\cT}{{\cal T}}
\newcommand{\cC}{{\cal C}}
\newcommand{\cQ}{{\cal Q}}

\begin{document}

\title[Giving up the ghost]{Giving up the ghost\footnote{Talk given by the
first author at the 5th International Symposium on Quantum Theory and
Symmetries (QTS5) held at the University of Vallodolid, Spain, July 2007.}}

\author{Carl~M~Bender${}^\dag$ and Philip~D~Mannheim${}^\ddag$}

\address{${}^\dag$Physics Department, Washington University, St. Louis, MO
63130, USA\\{\tt electronic address: cmb@wustl.edu}}

\address{${}^\ddag$Department of Physics, University of Connecticut, Storrs, CT
06269, USA\\{\tt electronic address: philip.mannheim@uconn.edu}}

\date{today}

\begin{abstract}
The Pais-Uhlenbeck model is a quantum theory described by a higher-derivative
field equation. It has been believed for many years that this model possesses
ghost states (quantum states of negative norm) and therefore that this model
is a physically unacceptable quantum theory. The existence of such ghost states
was believed to be attributable to the field equation having more than two
derivatives. This paper shows that the Pais-Uhlenbeck model does not possess any
ghost states at all and that it is a perfectly acceptable quantum theory. The
supposed ghost states in this model arise if the Hamiltonian of the model is
(incorrectly) treated as being Dirac Hermitian (invariant under combined matrix
transposition and complex conjugation). However, the Hamiltonian is {\it not}
Dirac Hermitian, but rather it is $\cP\cT$ symmetric. When it is quantized
correctly according to the rules of $\cP\cT$ quantum mechanics, the energy
spectrum is real and bounded below and all of the quantum states have positive
norm.
\end{abstract}

\pacs{11.30.Er, 12.38.Bx, 2.30.Mv}
\submitto{\JPA}

\section{Introduction}
\label{s1}

{\it Ghost} states are quantum states having negative norm. If a quantum theory
has ghost states, it is fundamentally unacceptable because the norm of a quantum
state is interpreted as a probability, and a negative probability is forbidden
on physical grounds.

One can present a simple and apparently compelling argument that if the field
equation of a quantum theory is a differential equation of higher than second
order in the time derivative, the theory must have ghost states. However, as we
will show in this paper, this argument is incorrect.

The argument goes as follows: Suppose, for example, that a field equation is
fourth order in the time derivative. If we take the Fourier transform of this
equation, we obtain the corresponding propagator $G(E)$ whose denominator is a
fourth-order polynomial in the energy $E$. In factored form such a propagator in
Euclidean space would have the form
\begin{equation}
G(E)=\frac{1}{(E^2+m_1^2)(E^2+m_2^2)}.
\label{e1}
\end{equation}
Observe that $G(E)$ describes the propagation of two kinds of states, one of
mass $m_1$ and the other of mass $m_2$. Assuming without loss of generality that
$m_2>m_1$, we can rewrite this propagator in the form of a partial fraction 
\begin{equation}
G(E)=\frac{1}{m_2^2-m_1^2}\left(\frac{1}{E^2+m_1^2}-\frac{1}{E^2+m_2^2}\right).
\label{e2}
\end{equation}
Evidently, $m_2$ is a state of {\it negative} probability because its residue
contribution to the propagator is negative. This appears to contradict the well
known form of the Lehmann representation: Recall that when the two-point Green's
function is expressed in the Lehmann representation, the condition that all
quantum states have positive norm implies that the residues of all intermediate
propagating states must be strictly positive \cite{BARTON}.

However, this argument is {\it incorrect} because it contains an implicit
assumption; namely, that the inner product for the Hilbert space of quantum
states is the Dirac inner product. The Dirac adjoint, which is indicated by the
notation ${}^\dag$, consists of combined matrix transposition and complex
conjugation. If another inner product is used, such as the inner product that
arises in $\cP\cT$ quantum mechanics \cite{REV}, then the negative sign in the
coefficient of the $1/(E^2+m_2^2)$ term in (\ref{e2}) does not necessarily
indicate the presence of a ghost state.

In $\cP\cT$ quantum mechanics the Hamiltonian $H$ is not Dirac Hermitian, $H\neq
H^\dag$, but is instead invariant under the more physical discrete symmetry of
space-time reflection $H=H^{\cP\cT}$ \cite{REV,BB,DDT}. Here, parity $\cP$ is a
linear operator that performs space reflection and $\cT$ is an antilinear
operator that performs time reversal. If the energy spectrum of $H$ is real and
positive, then the $\cP \cT$ symmetry of $H$ is unbroken, and it can be shown
that $H$ possesses a hidden reflection symmetry represented by the linear
operator $\cC$. Because $\cC$ is a reflection symmetry, it satisfies $\cC^2=1$.
The $\cC$ operator commutes with the Hamiltonian, $[H,\cC]=0$, and also with the
$\cP\cT$ operator, $[\cC,\cP\cT]=0$ \cite{BBJ}. In terms of $\cC$ one can then
construct a new inner product for which the adjoint is the combined $\cC\cP\cT$
operation: $\langle|\equiv|\rangle^{\cC\cP\cT}$. Using this new inner product,
the norm of a state is strictly positive \cite{BBJ,Gey,AM-metric}. Thus, given a
$\cP\cT$-symmetric Hamiltonian that is not Hermitian in the Dirac sense, one is
not free to choose the inner product {\it ab initio} because the inner product
is dynamically determined by the Hamiltonian itself. Time evolution in $\cP\cT$
quantum mechanics is unitary because $\cC\cP\cT$ commutes with $H$ in the
time-evolution operator $e^{-iHt}$.

In this paper we demonstrate that the famous Pais-Uhlenbeck model \cite{PU},
which has a higher-derivative field equation, is defined by a non-Hermitian $\cP
\cT$-symmetric Hamiltonian. We show that the long-held belief that this model
has ghost states is in fact not correct, and we do so by calculating the $\cC$
operator exactly and in closed form, thereby identifying the inner product that
is consistent with the Pais-Uhlenbeck Hamiltonian.

\section{Pais-Uhlenbeck Model}
\label{s2}

The Pais-Uhlenbeck oscillator model is defined by the higher-derivative,
acceleration-dependent action
\begin{equation}
I=\frac{\gamma}{2}\int dt\left[{\ddot z}^2-\left(\omega_1^2
+\omega_2^2\right){\dot z}^2+\omega_1^2\omega_2^2z^2\right],
\label{e3}
\end{equation}
where $\gamma$, $\omega_1$, and $\omega_2$ are all positive constants. We may
assume without loss of generality that $\omega_1\geq\omega_2$. If we vary the
action, we obtain the {\it fourth-order} field equation
\begin{equation}
z^{''''}(t)+(\omega_1^2+\omega_2^2)z^{''}(t)+\omega_1^2\omega_2^2z(t)=0.
\label{e4}
\end{equation}

One can construct the Pais-Uhlenbeck Hamiltonian by introducing a new dynamical
variable $x$ and obtaining a description of the theory using two degrees of
freedom \cite{MANN}:
\begin{equation}
H=\frac{p_x^2}{2\gamma}+p_zx+\frac{\gamma}{2}\left(\omega_1^2+\omega_2^2
\right)x^2-\frac{\gamma}{2}\omega_1^2\omega_2^2z^2.
\label{e5}
\end{equation}
In the Fock-space representation there are now two sets of creation and
annihilation operators:
\begin{eqnarray}
z&=&a_1+a_1^\dagger+a_2+a_2^\dagger,\nonumber\\
p_z&=&i\gamma\omega_1\omega_2^2
(a_1-a_1^\dagger)+i\gamma\omega_1^2\omega_2(a_2-a_2^\dagger),\nonumber\\
x&=&-i\omega_1(a_1-a_1^\dagger)-i\omega_2(a_2-a_2^\dagger),\nonumber\\
p_x&=&-\gamma\omega_1^2 (a_1+a_1^\dagger)-\gamma\omega_2^2(a_2+a_2^\dagger).
\label{e6}
\end{eqnarray}
These operators satisfy the usual set of commutation relations, with the
nonzero commutator given by
\begin{equation}
\omega_1[a_1,a_1^\dag]=-\omega_2[a_2,a_2^\dag]=\frac{1}{2\gamma
\left(\omega_1^2-\omega_2^2\right)}.
\label{e7}
\end{equation}
In terms of these operators the Hamiltonian for the Pais-Uhlenbeck model is
\begin{equation}
H=2\gamma(\omega_1^2-\omega_2^2)(\omega_1^2 a_1^\dag a_1-\omega_2^2a_2^\dag a_2)
+\half(\omega_1+\omega_2).
\label{e8}
\end{equation}

There appear to be only two possible realizations of the commutation relations
in (\ref{e7}), and we enumerate these below:

\begin{itemize}
\item[(I)] If $a_1$ and $a_2$ annihilate the 0-particle state $|\Omega\rangle$,
\begin{equation}
a_1|\Omega\rangle=0, \qquad a_2|\Omega\rangle=0,
\label{e9}
\end{equation}
then the energy spectrum is real and bounded below. The state $|\Omega\rangle$
is the ground state of the theory and it has zero-point energy $\half\left(
\omega_1+\omega_2\right)$. The problem with this realization is that the excited
state $a_2^\dag|\Omega\rangle$, whose energy is $\omega_2$ above ground state,
has a {\it negative Dirac norm} given by $\langle\Omega|a_2a_2^\dag|\Omega
\rangle$.

\item[(II)] If $a_1$ and $a_2^\dag$ annihilate the 0-particle state $|\Omega
\rangle$,
\begin{equation}
a_1|\Omega\rangle=0,\qquad a_2^\dag|\Omega\rangle=0,
\label{e10}
\end{equation}
then the theory is free of negative-norm states. However, this realization has a
different and equally serious problem; namely, that the energy spectrum is
unbounded below.

\end{itemize}

The two realizations (I) and (II) above are clearly unacceptable physically and
they characterize the generic problems that are thought to plague
higher-derivative quantum-mechanical theories and quantum field theories.
The Pais-Uhlenbeck model has been believed to be unphysical because there appear
to be no other realizations for which both the energy spectra and the norms
of the states are positive.

We emphasize that there may be many realizations of a given Hamiltonian
depending on the boundary conditions that are imposed on the coordinate-space
eigenfunctions. Let us consider two elementary examples:

\subsection{Harmonic oscillator $H=p^2+x^2$}

This Hamiltonian seems to be positive-definite because it is a sum of squares,
and one might therefore expect that the spectrum of this Hamiltonian would be
positive. However, this is incorrect. There are actually two distinct possible
realizations, that is, solutions to the Schr\"odinger equation eigenvalue
problem
\begin{equation}
-\psi''(x)+x^2\psi(x)=E\psi(x),
\label{e11}
\end{equation}
and these realizations are distinguished by the boundary conditions that are
imposed on the eigenfunction $\psi(x)$. 

If we require that $\psi(x)$ vanish exponentially fast as $|x|\to\infty$ with
$\rm{arg}\,x$ inside two $90^\circ$ Stokes wedges \cite{BO} centered about the
positive- and negative-real axes, then the spectrum is strictly positive and the
$n$th eigenvalue is given exactly by $E_n=2n+1$ ($n=0,\,1,\,2,\,3,\,\ldots$). On
the other hand, if we require that $\psi(x)$ vanish exponentially fast as $|x|
\to\infty$ with $\rm{arg}\,x$ inside two $90^\circ$ Stokes wedges centered about
the positive- and negative-{\it imaginary} axes, then the spectrum is strictly
negative and the $n$th eigenvalue is given exactly by $E_n=-2n-1$ ($n=0,\,1,\,2,
\,3,\,\ldots)$ \cite{REV}.

\subsection{Anharmonic oscillator $H=p^2-x^4$ with an ``upside-down'' potential}

This oscillator has many possible realizations. For example, if we do not
require that the solution to the Schr\"odinger equation 
\begin{equation}
-\psi''(x)-x^4\psi(x)=E\psi(x),
\label{e12}
\end{equation}
satisfy any boundary conditions at all on the real-$x$ axis, then the spectrum
is continuous and unbounded below. 

On the other hand, if we require that $\psi(x)$ vanish exponentially fast in a 
pair of $60^\circ$ Stokes wedges centered about the rays ${\rm arg}\,x=-
30^\circ$ and ${\rm arg}\,x=-150^\circ$, then the spectrum is real, discrete,
and strictly positive, and in fact the spectra of $H=p^2-x^4$ and of $\tilde H=
p^2+4x^4-2\hbar\,x$, where $\hbar$ is Planck's constant, are exactly identical
\cite{QUAR}. In $\tilde H$ the term proportional to $\hbar$ vanishes in the
classical limit and is thus a quantum anomaly \cite{QUAR}.

\section{Pais-Uhlenbeck model as a $\cP\cT$ quantum theory}
\label{s3}

To make sense of the Pais-Uhlenbeck model as a consistent and physical quantum 
theory, it is necessary to interpret the Pais-Uhlenbeck Hamiltonian (\ref{e5})
as a $\cP\cT$ quantum theory. The detailed analysis is given in Ref.~\cite{BM},
and we summarize it here.

We begin by modifying $H$ in (\ref{e5}) by making the substitution $y=-iz$ (and
the corresponding substitution $q=ip_z$ to enforce $[y,q]=i$) to obtain the
modified Hamiltonian 
\begin{equation}
H=\frac{p^2}{2\gamma}-iqx+\frac{\gamma}{2}\left(\omega_1^2+\omega_2^2
\right)x^2+\frac{\gamma}{2}\omega_1^2\omega_2^2y^2.
\label{e13}
\end{equation}
Here, we have simplified the notation by replacing $p_x$ by $p$. The operators
$p$, $x$, $q$, and $y$ are now formally Hermitian, but because of the $-iqx$
term $H$ has become explicitly complex and is manifestly not Dirac Hermitian.
We stress that this non-Hermiticity property is not apparent in the original
form of the Pais-Uhlenbeck Hamiltonian. This surprising and unexpected emergence
of a non-Hermitian term in the Pais-Uhlenbeck Hamiltonian provides insight into
the origin of the infamous ghost problem of the Pais-Uhlenbeck model.

Next, we make an unusual assignment for the properties of the dynamical
variables under space and time reflection: We take $p$ and $x$ to transform like
conventional coordinate and momentum variables under $\cP$ and $\cT$ reflection,
but we define $q$ and $y$ to transform {\it unconventionally} in a way not seen
in previous studies of $\cP\cT$ quantum mechanics \cite{REV}. (In the language
of quantum field theory, $q$ and $y$ transform as parity {\it scalars} instead
of {\it pseudoscalars}.) Note that $q$ and $y$ also have abnormal behavior under
time reversal. We summarize the symmetry properties of these operators in the
following table:
\begin{equation}
\begin{array}{c|cccc}
&p&x&q&y\\
\hline
\cP&-&-&+&+\\
\cT&-&+&+&-\\
\cP\cT&+&-&+&-\\
\end{array}
\label{e14}
\end{equation}
Under these definitions the Pais-Uhlenbeck Hamiltonian (\ref{e13}) is $\cP\cT$
symmetric. Also, the spectrum in the realization (I) in Sec.~\ref{s2} is
entirely real, so the $\cP\cT$ symmetry of the Pais-Uhlenbeck Hamiltonian is
unbroken.

The next step is to calculate the hidden symmetry operator $\cC$. As discussed
Sec.~\ref{s1}, the operator $\cC$ satisfies a system of three algebraic
equations:
\begin{equation}
\cC^2=1,\quad[\cC,\cP\cT]=0,\quad [\cC,H]=0.
\label{e15}
\end{equation}
In previous investigations \cite{PREV} it was found that $\cC$ has general form
\begin{equation}
\cC=e^{\cQ}\cP,
\label{e16}
\end{equation}
where $\cQ$ is a real function of dynamical variables and it is Hermitian in the
Dirac sense. In past studies it was found that $\cQ$ was odd in the momentum
variables and even in coordinate variables. However, because of the abnormal
behaviors of the $y$ and $q$ operators in (\ref{e14}), the exact solution to the
three simultaneous algebraic equations for $\cC$ in (\ref{e15}) exhibits an
unusual and previously unobserved structure for $\cQ$:
\begin{equation}
\cQ=\alpha pq+\beta xy,
\label{e17}
\end{equation}
where
\begin{eqnarray}
\beta&=&\gamma^2\omega_1^2\omega_2^2\alpha\quad{\rm and}\quad
\sinh(\sqrt{\alpha\beta})=\frac{2\omega_1\omega_2}{\omega_1^2-\omega_2^2}.
\label{e18}
\end{eqnarray}

Once the $\cQ$ operator is known, we can use it to find a Hamiltonian $\tilde H$
that is Hermitian in the Dirac sense by means of the similarity transformation
\cite{MMMM}
\begin{equation}
\tilde{H}=e^{-\cQ/2}He^{\cQ/2}.
\label{e19}
\end{equation}
When we perform this transformation using the operator $\cQ$ in (\ref{e17}), we
find that
\begin{equation}
{\tilde H}=e^{-\cQ/2}He^{\cQ/2}=\frac{p^2}{2\gamma}+\frac{q^2}{2\gamma\omega_1^2
}+\frac{\gamma}{2}\omega_1^2x^2+\frac{\gamma}{2}\omega_1^2\omega_2^2y^2.
\label{e20}
\end{equation}
Observe that the spectrum of ${\tilde H}$ is manifestly real because this
Hamiltonian is the sum of two harmonic-oscillator Hamiltonians. Furthermore, the
spectrum is positive because the transformations we have used to obtain $\tilde
H$ ensure that its eigenfunctions vanish in wedges containing the real axis. But
${\tilde H}$ is related to the original Pais-Uhlenbeck Hamiltonian by the
isospectral similarity transformation in (\ref{e19}). Thus, despite the $-iqx$
term in (\ref{e13}), the positivity of the Pais-Uhlenbeck Hamiltonian is proved.

To see why there are no ghosts and that the norm associated with the
Pais-Uhlenbeck model is strictly positive, we note that the eigenstates
$|{\tilde n}\rangle$ of ${\tilde H}$ have a positive inner product and they are
normalized in the standard Dirac way using the inner product $\langle{\tilde n}|
{\tilde n}\rangle=1$, where the bra vector is the Dirac-Hermitian adjoint
${}^\dag$ of the ket vector. Equivalently, for the eigenstates $|n\rangle$ of
the Hamiltonian $H$, because the vectors are mapped by $|{\tilde n}\rangle=
e^{-\cQ/2}|n\rangle$, the eigenstates of $H$ are normalized as
\begin{equation}
\langle n|e^{-\cQ}|m\rangle=\delta(m,n)\quad{\rm with}\quad\sum|n\rangle\langle
n|e^{-\cQ}=1.
\label{e21}
\end{equation}
The norm in (\ref{e21}) is relevant for the Pais-Uhlenbeck model, with $\langle
n|e^{-\cQ}$ rather than $\langle n|$ being the appropriate conjugate for $|n
\rangle$. Since the norm in (\ref{e21}) is positive and because $[H,\cC\cP\cT]=
0$, the Pais-Uhlenbeck Hamiltonian $H$ generates unitary time evolution.

\section{Discussion}

The Pais-Uhlenbeck model is not the first instance for which it can be shown
that what was previously thought to be a ghost is actually a conventional
quantum state having a positive norm. The first model that was discovered to
have a false ghost state was the Lee model. The Lee model was proposed in 1954
as a quantum field theory in which mass, wave-function, and charge
renormalization could be performed exactly and in closed form \cite{L1}.
However, in 1955 K\"all\'en and Pauli showed that when the renormalized coupling
constant is larger than a critical value, the Hamiltonian becomes non-Hermitian
(in the Dirac sense) and a ghost state appears \cite{L2}. The appearance of the
ghost was assumed to be a fundamental defect of the Lee model. However, the
non-Hermitian Lee-model Hamiltonian is $\cP\cT$ symmetric. When the norms of the
states of this model are determined using the $\cC$ operator, which, as in the
Pais-Uhlenbeck model, can be calculated in closed form, the ghost state is seen
to be an ordinary physical state having positive norm \cite{LEE}. Thus, the
following words by Barton \cite{BARTON} are {\it not true}: ``A non-Hermitian
Hamiltonian is unacceptable partly because it may lead to complex energy
eigenvalues, but chiefly because it implies a non-unitary S matrix, which fails
to conserve probability and makes a hash of the physical interpretation.''

Thus, there are now {\it two} independent models in which one can show that what
was believed to be a ghost state is actually an ordinary state having positive
norm. This suggests that there may be many more examples of quantum theories
that have been abandoned as being unphysical and that can be repaired by using
the methods of $\cP\cT$ quantum mechanics. The problem of ghosts arises in
quantizing gravity, and we hope that the methods of $\cP\cT$ quantum mechanics
will be able to establish that the classical theory of gravity can be
consistently quantized without the appearance of ghosts.

\vspace{0.5cm}

\footnotesize
\noindent
CMB is supported by a grant from the U.S. Department of Energy.

\end{document}